# Non-Singular Magnetic Monopoles

**S Khademi[1], M Shahsavari[2] and A M Saeed[1]**

[1] Department of Physics, Zanjan University, 5th Km of Tabriz Road, Zanjan, Iran

[2] Institute of Geophysics, Tehran University, Amir-Abad Street, Tehran, Iran

Email:skhademi@mail.znu.ac.ir and siamakkhademi@yahoo.com

**Abstract.** Magnetic Monopole is a consequence of the existence of the duality symmetry in electrodynamics. Although, no conclusive experimental evidences have so far been found but the subject is still of much interest to physicists. The theory of magnetic monopoles was first proposed by Dirac in 1931 and soon after it was studied by physicist of many disciplines specially particle physics, quantum filed theory and the non-linear Soliton equations. One important consequence of the magnetic monopole theory is the quantization of the electric charge which was first derived by Dirac. In the definition of the classical magnetic monopole, the concept of Dirac string is used. Dirac string is the locus of the points where the vector potential is not well-behaved. On the other hand by introducing the idea of magnetic monopoles, Maxwell's equations became symmetrical with respect to the magnetic and electric fields, but they still remain unsymmetrical as far as the scalar and the vector potentials are concerned. In this work the electric and magnetic fields are redefined in terms of some new scalar and vector potentials, respectively, as a result of which the Maxwell's equations become symmetrical with respect to the potentials too. One advantage of using this formulation is that one can discard the Dirac string all together. Finally, definition of the new potentials guarantees the Lorentz invariance of equations.

## 1. Introduction

The concept of symmetry plays a central role in Physics and other fields of science [1]. Moreover educationally, it is essential for students to know the various aspects of the phenomena. Symmetry, apart from its aesthetic aspects is a powerful tool in predicting and forecasting physical laws and effects, something which makes it all more important for students to learn. The application of symmetry in predicting new particles is a well known technique e.g. Lorentz invariance of the quantum mechanical equations of particles led to the prediction of anti-particles in 1930. In 1993 anti-electrons (positrons) were discovered by C. D. Anderson [2]. Magnetic monopole is also an important fundamental particle whose existence is postulated based on the duality symmetry. Although so far no experiments have revealed such particles but some physicists are hoping that these particles can experimentally be detected in future.

The quantization of the electric charges is an important consequence of the investigation of the dual symmetry in quantum electrodynamics, by Dirac [3]. In his work the vector potential was singular and he considered the concept of Dirac string in the formulation [4]. In this work the duality symmetry is re-examined and it is shown that the singularities originate from the lack of the dual symmetry in the definition of the electromagnetic fields in terms of potentials. Symmetrization of the Maxwell's equations with regards to the potentials would result in the definition of new scalar and vector potentials.

## 2. Dual Symmetry of Maxwell's equations with respect to electromagnetic fields

The well-known Maxwell's equations consist of two isotropic and two non-isotropic equations

$$\nabla \cdot \mathbf{E} = 4\pi\rho_e, \qquad \nabla \cdot \mathbf{B} = 0,$$
$$\nabla \times \mathbf{B} - \frac{1}{c}\frac{\partial \mathbf{E}}{\partial t} = \frac{4\pi}{c}\mathbf{j}_e, \qquad \nabla \times \mathbf{E} + \frac{1}{c}\frac{\partial \mathbf{B}}{\partial t} = 0. \tag{1}$$

# Non-Singular Magnetic Monopoles

If the magnetic monopoles exist, in exactly the same way as that of the electric charges, then the magnetic field due to such monopoles is given by

$$\mathbf{B} = q_m \frac{\mathbf{r}}{r^3}. \tag{2}$$

In Eq. (2) $q_m$ is the magnetic charge. The sign of $N$ and $S$ magnetic charges would differ as

$$sign(q_N) = -sign(q_S). \tag{3}$$

With the assumption of the existence of the magnetic monopoles the Maxwell's equations would then have a symmetrical form [5]

$$\nabla \cdot \mathbf{E} = 4\pi\rho_e, \qquad \nabla \cdot \mathbf{B} = 4\pi\rho_m,$$
$$\nabla \times \mathbf{B} - \frac{1}{c}\frac{\partial \mathbf{E}}{\partial t} = \frac{4\pi}{c}\mathbf{j}_e, \quad \nabla \times \mathbf{E} + \frac{1}{c}\frac{\partial \mathbf{B}}{\partial t} = -\frac{4\pi}{c}\mathbf{j}_m. \tag{4}$$

and they would also be invariant under the following transformations

$$\mathbf{j}_m \to -\mathbf{j}_e, \qquad \mathbf{E} \to \mathbf{B},$$
$$\rho_e \to \rho_m, \qquad \mathbf{B} \to -\mathbf{E}, \tag{5}$$
$$\rho_m \to -\rho, \qquad \mathbf{j}_e \to \mathbf{j}_m,$$

which are called dual transformation [5]. If one expresses the electromagnetic fields in terms of the scalar and vector potentials by

$$\mathbf{B} = \nabla \times \mathbf{A},$$
$$\mathbf{E} = -\frac{1}{c}\frac{\partial \mathbf{A}}{\partial t} - \nabla\varphi, \tag{6}$$

then the Gauss' law for the electric charges would no longer hold because

$$\nabla \cdot \mathbf{B} = \nabla \cdot (\nabla \times \mathbf{A}) = 0 \neq 4\pi\rho_m. \tag{7}$$

To resolve this, Dirac used a singular vector potential. He first considered the integral form of the Gauss' law for a magnetic field over an open surface. A Gaussian open surface is defined as a sphere which has a hole of radius $R$ cut out of it (see Fig. 1)

$$\int_{\substack{S-Open \\ R \to 0}} \mathbf{B} \cdot d\mathbf{s} = 4\pi q_m. \tag{8}$$

Now, using the Eq. (6) and the Stock's law, Eq. (8) is transformed into a line integral around the cut out area of radius $R$,

$$\int_{\substack{S-Open \\ R \to 0}} \mathbf{B}.d\mathbf{s} = \int_{\substack{S \\ R \to 0}} (\nabla \times \mathbf{A}).d\mathbf{s} = \int_{\substack{C \\ R \to 0}} \mathbf{A}.d\mathbf{l} = 4\pi\rho_m. \tag{9}$$

## Non-Singular Magnetic Monopoles

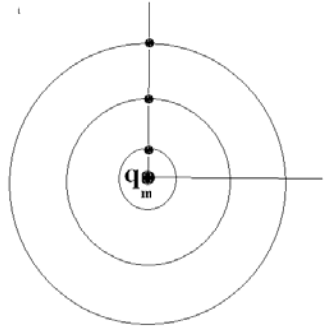
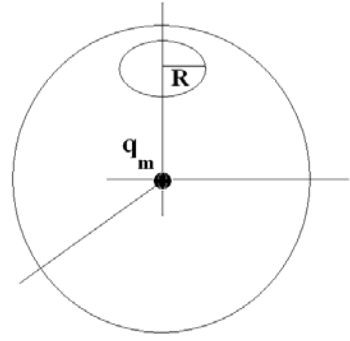

Figure 2: The bold points are the locus of the singularities on the spherical surfaces around the magnetic charge $q_m$.

Figure 1: The (Gaussian) surface, with a hole of radius R, around a magnetic charge $q_m$.

Now the radius R is let to tend to zero till a singular point is obtained on the surface. In the obtained singular point the vector potential is not well-defined. As an example, in a convenient gauge or with a suitable choice of a vector potential, e.g.,

$$\mathbf{A} = q_m \frac{1 + Cos\theta}{rSin\theta} \hat{\varphi}, \tag{10}$$

the singularity would end up on the Z-axis ($\theta = 0$). However one can find many Gaussian spherical surfaces which would have the magnetic charge in their center and a singular point on their surface. The locus of such singular points would form a line or string which is known as Dirac string. In another word in this method a singular string is attached to each magnetic monopole, which stretches from the monopole to infinity.

Unlike electromagnetic fields, potentials are not measurable quantities and in spite of having singularities in the potentials the electromagnetic fields are not singular themselves. However the Hamiltonian of an interacting charged particle with external electromagnetic fields explicitly depends on the electromagnetic potentials

$$H = \frac{1}{2m}(\mathbf{P} - \frac{q}{c}\mathbf{A})^2 + q\varphi, \tag{11}$$

therefore, in quantum electrodynamics the Schrödinger's equation for a charged particle interacting with the electromagnetic fields of a magnetic monopole would be a singular differential equation. On the other hand, using the definition of electromagnetic fields in terms of ordinary potentials, Eq. (6) results in non-symmetrical Maxwell's equations. In another word the Maxwell's equations do not have the duality symmetry with respect to the ordinary potentials.

### 3. Dual Symmetry of Maxwell's equations with respect to the electromagnetic potentials

In order to symmetrize the Maxwell's equations, two new potentials, a vector potential $G$ and a scalar potential $\psi$, in addition to the ordinary potentials are introduced such that

$$\mathbf{B} = \frac{1}{c}\frac{\partial \mathbf{G}}{\partial t} + \nabla \psi + \nabla \times \mathbf{A},$$

$$\mathbf{E} = -\frac{1}{c}\frac{\partial \mathbf{A}}{\partial t} - \nabla \varphi + \nabla \times \mathbf{G}. \tag{12}$$

Equations (12) are invariant under the dual transformation

## Non-Singular Magnetic Monopoles

$$\varphi \to -\psi, \quad \mathbf{A} \to -\mathbf{G}, \quad \mathbf{E} \to \mathbf{B},$$
$$\psi \to \varphi, \quad \mathbf{G} \to \mathbf{A}, \quad \mathbf{B} \to -\mathbf{E}.$$
(13)

Similarly, Eqs. (12) are invariant under the Gauge transformations

$$\mathbf{G} \to \mathbf{G} + \nabla g, \quad \mathbf{A} \to \mathbf{A} + \nabla f,$$
$$\psi \to \psi - \frac{1}{c}\frac{\partial g}{\partial t}, \quad \varphi \to \varphi - \frac{1}{c}\frac{\partial f}{\partial t},$$
(14)

too. In these transformations, in addition to the choice of the well-behaved function $f$, the choice of function $g$ is also arbitrary. The gauge invariance of the formulation in terms of the two arbitrary functions $f$ and $g$ results in the conservation of electric and magnetic charges, respectively. The Lorentze covariant form of Eq. (12) is obtained as

$$F_{\mu\nu} = \partial_\mu A_\nu - \partial_\nu A_\mu + \tfrac{1}{2}\varepsilon_{\mu\nu\alpha\beta}(\partial^\alpha G^\beta - \partial^\beta G^\alpha).$$
(15)

In Eq. (15), the field strength tensor, $F_{\mu\nu}$ is defined in terms of $A_\mu \equiv (\varphi, -\mathbf{A})$ and $G^\nu \equiv (\psi, \mathbf{G})$. Also $\varepsilon_{\mu\nu\alpha\beta}$ is fourth order Levi-Civita tensor. Now considering the new potentials in definition of electromagnetic fields, Eq.(12), one may obtain the divergence of the magnetic field which is no longer zero and is given by

$$\nabla \cdot \mathbf{B} = \nabla \cdot (\tfrac{1}{c}\frac{\partial \mathbf{G}}{\partial t} + \nabla \psi + \nabla \times \mathbf{A}) = \tfrac{1}{c}\frac{\partial(\nabla \cdot \mathbf{G})}{\partial t} + \nabla^2 \psi = \rho_m.$$
(16)

With an special gauge, $\nabla \cdot \mathbf{G} = 0$, the Poisson equation for the potential $\psi$ is obtained as

$$\nabla^2 \psi = \rho_m.$$
(17)

The solutions of the Poisson equation (17) for a magnetic monopole with charge $q_m$, is then given by

$$\psi = \frac{q_m}{r}.$$
(18)

The magnetic field is given by the gradient of the scalar potential $\psi$

$$\mathbf{B} = \nabla \psi.$$
(19)

Evidently, Eqs. (18) and (19) lead to Eq. (2).

From the symmetric Maxwell's equations one may conclude that a moving electric charge particle gives rise to an electric field and symmetrically, a moving magnetic charged particle would also give rise to an electric field (see Figures 3 and 4). The electric and magnetic fields in Eq. (12) would now split in to longitudinal and transversal fields as given below

$$\mathbf{B}_L = \tfrac{1}{c}\frac{\partial \mathbf{G}_L}{\partial t} + \nabla \psi, \quad \mathbf{B}_T = \tfrac{1}{c}\frac{\partial \mathbf{G}_T}{\partial t} + \nabla \times \mathbf{A}_T,$$
$$\mathbf{E}_L = -\tfrac{1}{c}\frac{\partial \mathbf{A}_L}{\partial t} - \nabla \varphi, \quad \mathbf{E}_T = -\tfrac{1}{c}\frac{\partial \mathbf{A}_T}{\partial t} + \nabla \times \mathbf{G}_T.$$
(20)

In Eq. (20) the indices L and T denote the longitudinal and transverse components, respectively. The longitudinal electric and the transversal magnetic fields are produced by a moving electric charge, while the longitudinal magnetic and the transversal electric fields are produced by a moving magnetic charged particle. The duality symmetry (5) can clearly be seen in figures 3 and 4.

# Non-Singular Magnetic Monopoles

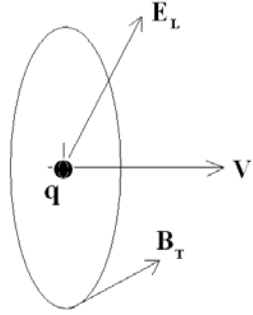
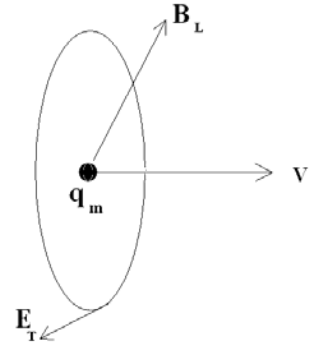

Figure 4: Longitudinal electric field $\mathbf{E_L}$ and transversal magnetic field $\mathbf{B_T}$ due to a moving electric charged particle $q$.

Figure 3: Transversal electric field $\mathbf{E_T}$ and longitudinal magnetic field $\mathbf{B_L}$ due to a moving magnetic charge $q_m$.

In studies concerning the magnetic as well as the electric charged particles, beside the gauge related to the ordinary potentials $\mathbf{A}$ and $\varphi$ one has an additional gauge related to the new magnetic potentials $\mathbf{G}$ and $\psi$. Therefore, in addition to the choice of a common gauge, the gauge related to the new potentials must also be specified. For example, the gauge $\nabla \cdot \mathbf{A} = 0$, $\nabla \cdot \mathbf{G} = 0$ which one may call it the Coulomb–Coulomb gauge and $\nabla \cdot \mathbf{G} = 0$, $\nabla \cdot \mathbf{A} + \frac{1}{c}\frac{\partial \varphi}{\partial t} = 0$ the Lorentz-Coulomb gauge. As it is self-explanatory, the first half of the gauge name is related to the electric potentials while the second half to the magnetic potentials.

## 4. Lorentz Force

One may divide the electric(magnetic) fields in to two parts: the field which is due to the electric charges, $\mathbf{E_e}$ ($\mathbf{B_e}$) and one which is due to magnetic charges, $\mathbf{E_m}$ ($\mathbf{B_m}$)

$$\mathbf{E} = -\frac{1}{c}\frac{\partial \mathbf{A}}{\partial t} - \nabla \varphi + \nabla \times \mathbf{G} = \mathbf{E}_e + \mathbf{E}_m,$$

$$\mathbf{B} = \frac{1}{c}\frac{\partial \mathbf{G}}{\partial t} + \nabla \psi + \nabla \times \mathbf{A} = \mathbf{B}_m + \mathbf{B}_e.$$

(21)

The subscripts e and m denote the fields due to the electric and magnetic charges respectively. In Eq. (21) the magnetic and electric fields due to a magnetic charge is given by $\mathbf{B}_m = \frac{1}{c}\frac{\partial \mathbf{G}}{\partial t} + \nabla \psi$ and $\mathbf{E}_m = \nabla \times \mathbf{G}$, respectively. The magnetic and electric fields resulting from an electric charge is also obtained from Eq. (6).

The force acting on an electric charge $q$ due to the electric and magnetic fields which are in turn due to an electric charge is given by

$$\mathbf{F_e} = q[\mathbf{E_e} + \tfrac{1}{c}\mathbf{V} \times \mathbf{B_e}].$$

(22)

The electromagnetic force, caused by a magnetic charge and acting on another magnetic charge $q_m$ is obtained by dual transformation of Eq. (22)

$$\mathbf{F}_m = q_m[\mathbf{B}_m - \tfrac{1}{c}\mathbf{V} \times \mathbf{E}_m].$$

(23)

In examination of the interaction between the magnetic charges and the electromagnetic fields created by electric charges and vice versa, it is important to note that there is no difference

# Non-Singular Magnetic Monopoles

between the electromagnetic field created by either types of charges and hence they are indistinguishable. As a result the Lorentz force acting on a magnetic or electric charge can respectively be written as

$$\mathbf{F_m} = q_m [(\frac{1}{c}\frac{\partial \mathbf{G}}{\partial t} + \nabla \psi + \nabla \times \mathbf{A}) - \frac{1}{c}\mathbf{V} \times (-\frac{1}{c}\frac{\partial \mathbf{A}}{\partial t} - \nabla \varphi + \nabla \times \mathbf{G})], \tag{24}$$

$$\mathbf{F_e} = q[(-\frac{1}{c}\frac{\partial \mathbf{A}}{\partial t} - \nabla \varphi + \nabla \times \mathbf{G}) + \frac{1}{c}\mathbf{V} \times (\frac{1}{c}\frac{\partial \mathbf{G}}{\partial t} + \nabla \psi + \nabla \times \mathbf{A})]. \tag{25}$$

Equations (24) and (25) are dual transformation of each other.

## 5. Conclusion

Equations (4), (24) and (25) form a complete set of classical equations which describe the dynamics of both the electric and magnetic charges as well as the electromagnetic fields. Therefore they could be used in tackling all forms of the classical electrodynamics problems in the presence of a magnetic charge. Also since there is no singularity involved in this formulation therefore there is no need for the Dirac string. This is a consequence of the introduction of the duality symmetry into the potentials. Equations (12) which give the new potentials, which are invariant under the duality transformation, remain invariant under Lorentz transformation too.